\documentclass[journal=jpclcd,manuscript=letter]{achemso}
\usepackage{amsmath,amssymb}
\usepackage{float}

\usepackage[T1]{fontenc}
\usepackage[utf8]{luainputenc}
\usepackage{float}
\usepackage{multirow}
\usepackage{lipsum}
\usepackage{mathrsfs}
\usepackage{setspace}
\usepackage{droidsans}
\usepackage{balance}
\usepackage{times,mathptmx}
\usepackage{braket}
\usepackage{sectsty}
\usepackage{graphicx}
\usepackage{amsmath,amssymb}
\usepackage{mathrsfs}
\usepackage{multirow} 
\usepackage{lastpage}
\usepackage{adjustbox}
\usepackage{upgreek}
\usepackage{color}
\usepackage{array}
\usepackage{float}
\usepackage{fancyhdr}
\usepackage{fnpos}
\usepackage{adjustbox}
\usepackage[english]{babel}
\DeclareUnicodeCharacter{0308}{~}
\usepackage[version=3]{mhchem} 

\author{Manjari Jain}
\email{Manjari.Jain@physics.iitd.ac.in [MJ]}
\author{Manish Kumar, Preeti Bhumla}
\author{Saswata Bhattacharya}
\email{saswata@physics.iitd.ac.in[SB]}
\phone{+91-11-2659 1359}
\affiliation[Indian Institute of Technology Delhi]
{Department of Physics, Indian Institute of Technology Delhi, New Delhi, India}
\title[An \textsf{achemso} demo]
{Lead Free Alloyed Double Perovskites: An Emerging Class of Materials from Many-Body Perturbation Theory}
\begin{document}

\begin{abstract}
	
The discovery of lead free all-inorganic alloyed double perovskites have revolutionized photovoltaic research, showing promising light emitting efficiency and its tunability. However, detailed studies regarding optical, exciton, polaron and transport properties remain unexplored. Here, we report a theoretical study on the variation of carrier-lattice interaction and optoelectronic properties of pristine as well as alloyed Cs$_2$AgInCl$_6$ double perovskites. We have employed many-body perturbation theory (G$_0$W$_0$@HSE06)  and density functional perturbation theory (DFPT) to compute exciton binding energy (E$_\textrm{B}$) and exciton lifetime of different alloyed double perovskites. We find that phonon scattering limits charge-carrier mobilities and thus, plays an important role in the development of high-efficiency perovskite photovoltaics. In view of this, dominant carrier-phonon scattering is observed via Fr\"{o}hlich mechanism near room temperature. Moreover, we observe a noticeable increase in hole and electron mobilities on alloying. We believe that our results will be helpful to gain a better understanding of the optoelectronic properties and lattice dynamics of these double perovskites.
  \begin{tocentry}
  \begin{figure}[H]%
 	\includegraphics[width=1.0\columnwidth,clip]{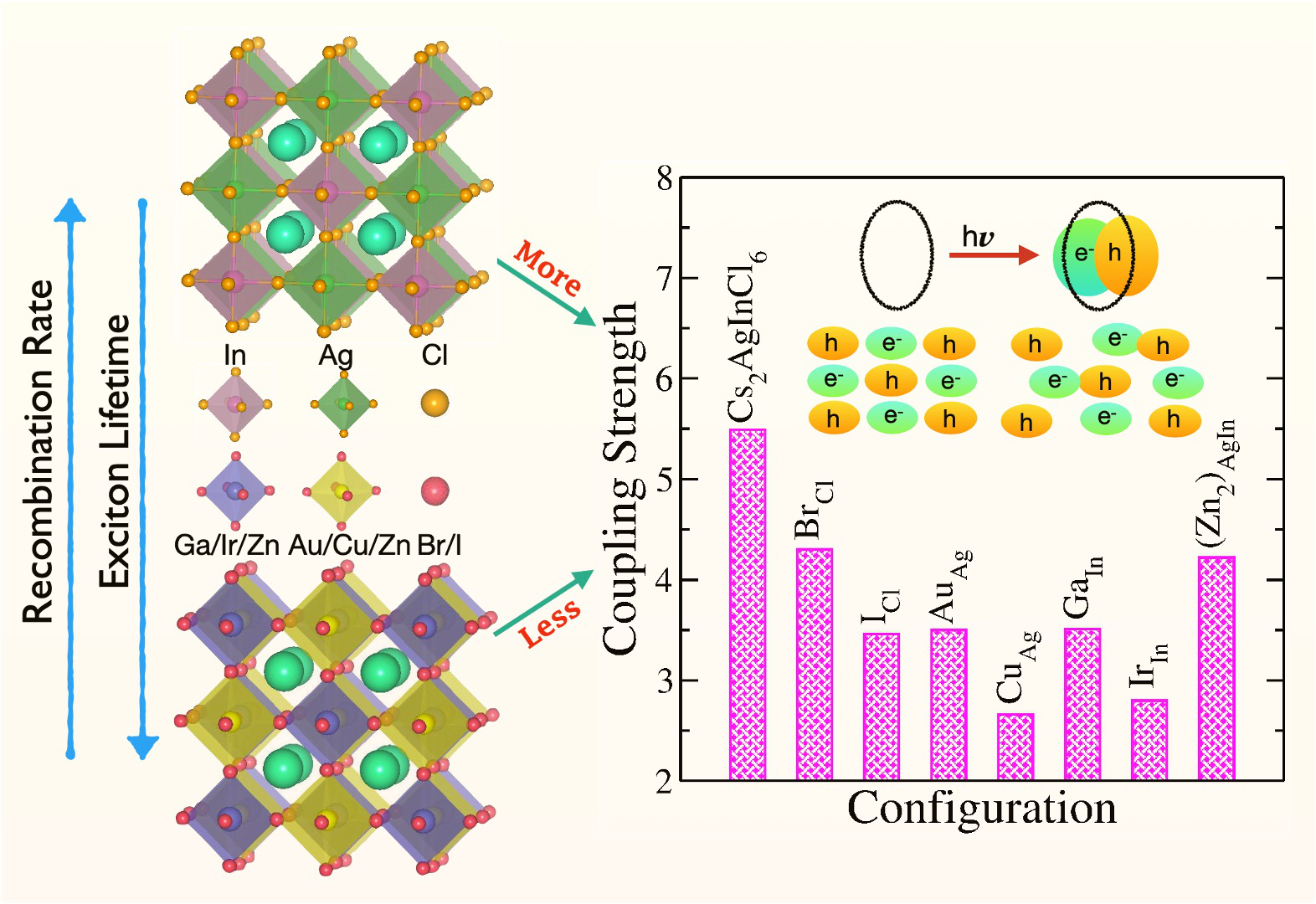}
  \end{figure}	
  \end{tocentry}
\end{abstract}
 In a remarkably short period of time, lead halide perovskites have emerged in an unprecedented way in the field of optoelectronics by virtue of their exceptional properties like suitable optical band gap, long carrier diffusion length, high carrier mobility and low manufacturing cost\cite{de2017solution,kovalenko2017properties,zhou2018metal,zhao2016organic,yang2018high,feng2018record,cho2017highly,tan2014bright,D0TC01484B}. The power conversion efficiency of lead halide perovskites-based solar cells has increased from 3.8\% to 25.5\% in the last decade\cite{doi:10.1021/ja809598r,nrel}. Despite their great potential, there are two major challenges including Pb toxicity and phase instability. The stability issues have been partly overcome by replacing organic cation with inorganic cesium ion~\cite{zhao2018high,xiang2018highly}. However, Pb toxicity is still a drawback and dealing with this issue without compromising the efficiency is of paramount importance.
 
 Over the past few years, lead free all-inorganic double perovskites with formula A$_2$B$'$B$''$X$_6$ have emerged as possible contenders due to low toxicity, intrinsic thermodynamic stability and small carrier effective mass~\cite{mcclure2016cs2agbix6,slavney2016bismuth,giustino2016lead,filip2018phase,schade2018structural,volonakis2017cs2inagcl6,filip2016band,zhang2018manipulation,D0TC02231D,C8TC03496F}. In particular, the double perovskite Cs$_2$AgInCl$_6$ has been the subject of several studies, revealing excellent environmental stability and promising optoelectronic properties. For example, Cs$_2$AgInCl$_6$ double perovskite exhibits a direct band gap as well as ultralong carrier lifetime (6 $\mu$s), which are suitable for photovoltaic applications\cite{https://doi.org/10.1002/adma.201706246,C7MH00239D}. Moreover, Luo \textit{et al.} reported that the alloying of Na cation in double perovskite Cs$_2$AgInCl$_6$ increases the photoluminescence efficiency by three orders and emits a warm white light with enhanced quantum efficiency~\cite{luo2018efficient}. Also, Cs$_2$AgInCl$_6$ has found to be successfully synthesized experimentally with the direct band gap of 3.3 eV and good stability~\cite{zhou2017composition}. However, the wide band gap of Cs$_2$AgInCl$_6$ doesn't show optical response in visible region. In order to reduce its band gap and expand the spectral response in visible light region, alloying with suitable elements could be the best solution. In one of our recent works, we have done the sublattice mixing by partial substitution of several metals M(I), M(II), M(III), and halogen X at Ag/In and Cl sites, respectively, to reduce the band gap of Cs$_2$AgInCl$_6$ and hence, enhancing its optical properties~\cite{doi:10.1063/5.0031336}. We have found that sublattices with Br and I substitutions at the Cl site, Cu(I) and Au(I) at the Ag site, Ga and Ir(III) at the In site, and Zn(II) at the Ag and In sites simultaneously have tuned the band gap in the visible region. In the present work, we have studied the excitonic properties and the role of electron-phonon coupling in these promising perovskites.
 
 Formation of excitons in optoelectronic materials greatly influences the charge separation properties. Hence, accurate estimation of excitonic parameters viz. exciton binding energy, exciton radius and exciton lifetime is of great importance in this class of materials. The excitons dissociate into free charge carriers, which effect the solar cell performance. Moreover, in order to explain multiple photophysical phenomena in perovskite materials, the concept of polarons has been used. The polaronic effect plays a crucial role in the excitaion dynamics and charge transport. The separation of free charge carriers is also influenced by the carrier mobility, which in turn depends on the electron-phonon (e-ph) coupling strength. Understanding the effect of electron-phonon coupling in terms of polaron mobility is important. In view of this, we attempt here to explore electronic and optical properties of Cs$_2$AgInCl$_6$ and its alloyed counterparts using state-of-the-art many-body perturbation theory. Various important aspects such as excitonic properties, carrier mobility, the role of electron-phonon coupling have been studied presumably for the first time in this class of materials. Using Fr\"{o}hlich model, we have also discussed the effect of electron-phonon coupling and calculated the polaron mobility.
 
\begin{figure}[b]
	\centering
	\includegraphics[width=0.8\textwidth]{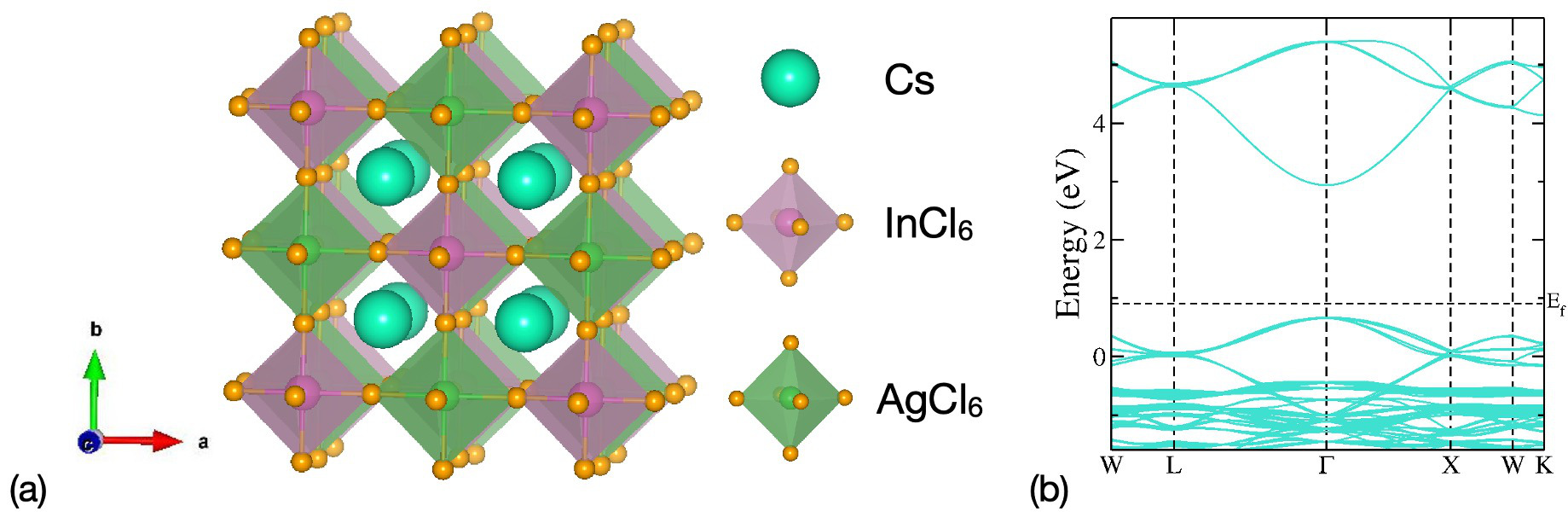}
	\caption{(a) Crystal structure of Cs$_2$AgInCl$_6$ double perovskite and (b) bandstructure of Cs$_2$AgInCl$_6$ using G$_0$W$_0$@HSE06. E$_\textrm{f}$ is fermi energy level.}
	\label{1}	
\end{figure}

According to Wannier-Mott model~\cite{waters2020semiclassical}, the exciton binding energy (E$_\textrm{B}$) for screened interacting electron-hole (e-h) pair is given by:
\begin{equation}
	\begin{split}
		\textrm{E}_\textrm{B}=\left(\frac{\mu}{\epsilon_\textrm{eff}^2}\right)\textrm{R}_\infty
		\label{eq1}
	\end{split}
\end{equation}
where, $\mu$ is the reduced mass, $\epsilon_\textrm{eff}$ is the effective dielectric constant and R$_\infty$ is the Rydberg constant.
In order to compute E$_\textrm{B}$, firstly we have calculated the effective mass of electrons and holes using Wannier-Mott approach by plotting E-$\textit{k}$ dispersion curve (see Fig.~\ref{1}(b) for pristine Cs$_2$AgInCl$_6$ and for different alloyed compounds, see section I of supplementary information (SI)). The conduction band minimium (CBm) and valence band maximum (VBM) are obtained at high symmetry point $\Gamma$ (0, 0, 0). We have done the parabolic fitting of the dispersion curves for calculating the effective mass of the electrons and holes. Further, we have taken the average of effective mass values from $\Gamma$ $\rightarrow$ L and $\Gamma$ $\rightarrow$ X directions. The effective mass can be calculated by using expression:
\begin{equation}
	\begin{split}
		\textrm{m}^*=\frac{\hbar^2}{\frac{d^2\textrm{E}(k)}{dk^2}}
		\label{eq1}
	\end{split}
\end{equation}
where m$^*$, E($\textit{k}$), $\textit{k}$, and $\hbar$ are the effective mass, energy, wave vector and reduced Planck's constant, respectively. The calculated effective mass and reduced mass in terms of rest mass of electron (m$_0$) for pristine Cs$_2$AgInCl$_6$ double perovskite and different alloyed compounds are listed in Table~\ref{1}. The obtained values of effective mass and reduced mass for Cs$_2$AgInCl$_6$ double perovskite are well in agreement with the previous findings~\cite{volonakis2016lead}.

\begin{table}[htbp]
	\caption{Effective mass of electron m$_\textrm{e}^\ast$, hole m$_\textrm{h}^\ast$ and reduced mass $\mu$ in terms of rest mass of electron m$_0$.} 
	\begin{center}
		\begin{tabular}[c]{|c|c|c|c|} \hline
			\textbf{Compounds} & \textbf{${\textrm{m}_\textrm{e}^*}$}  &  \textbf{${\textrm{m}_\textrm{h}^*}$} &  \textbf{${\mu}$}\\ \hline
			Cs$_2$AgInCl$_6$  & 0.29 & 1.07 & 0.23  \\ \hline
			Cs$_2$AgInBr$_{0.04}$Cl$_{5.96}$  & 0.28 & 0.80 & 0.21  \\ \hline
			Cs$_2$AgInI$_{0.04}$Cl$_{5.96}$   & 0.26 & 0.67 & 0.19  \\ \hline
			Cs$_2$Au$_{0.25}$Ag$_{0.75}$InCl$_6$   & 0.25 & 0.80 & 0.19  \\ \hline
			Cs$_2$Cu$_{0.25}$Ag$_{0.75}$InCl$_6$ & 0.27 & 1.09 & 0.22 \\ \hline
			Cs$_2$AgGa$_{0.25}$In$_{0.75}$Cl$_6$  & 0.27 & 0.84 & 0.21 \\ \hline
			Cs$_2$AgIr$_{0.25}$In$_{0.75}$Cl$_6$  & 0.29 & 0.86 & 0.22\\ \hline
			Cs$_2$Zn$_{0.50}$Ag$_{0.75}$In$_{0.75}$Cl$_6$ & 0.28 & 0.83 & 0.21  \\ \hline
		\end{tabular}
		\label{Table1}
	\end{center}
\end{table}
Now, our next task is to compute $\epsilon_\textrm{eff}$ in order to find the exciton binding energy of pristine Cs$_2$AgInCl$_6$ and different alloyed compounds. As it is reported earlier that the E$_\textrm{B}$ gets changed due to the lattice relaxation~\cite{freysoldt2014first} i.e., if $\omega_{\textrm{LO}}$ is the longitudinal optical phonon frequency and E$_\textrm{B}$$<<$$\hbar\omega_{\textrm{LO}}$, then one needs to consider lattice relaxation. Therefore, a value intermediate between the static electronic dielectric constant at high-frequency ($\epsilon_\textrm{$\infty$}$) and the static ionic dielectric constant at low frequency ($\epsilon_\textrm{static}$) should be considered for the $\epsilon_\textrm{eff}$. However, if E$_\textrm{B}$$>>$$\hbar\omega_{\textrm{LO}}$ then the ionic
contribution to the dielectric screening is negligible and, hence does not alter the E$_\textrm{B}$~\cite{bokdam2016role}. In such cases, $\epsilon_\textrm{eff}$ $\rightarrow$ $\epsilon_\textrm{$\infty$}$, where $\epsilon_\textrm{$\infty$}$ is the static value of dielectric constant at high frequency that mainly consists of electronic contribution.
\begin{figure}[h]
	\centering
	\includegraphics[width=0.8\textwidth]{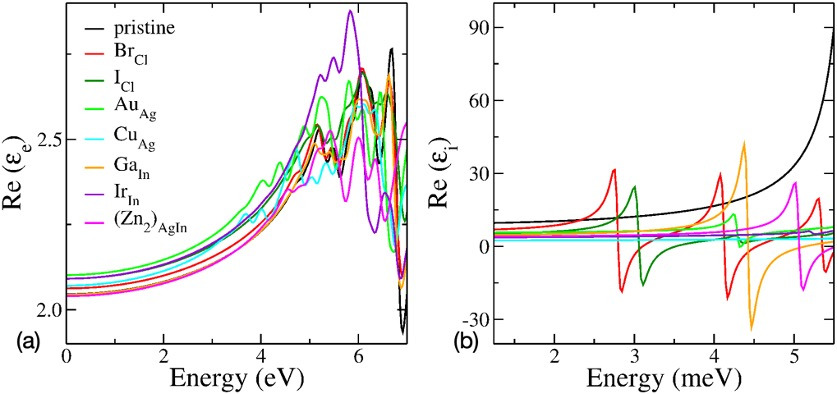}
	\caption{(a) Electronic contribution to the dielectric function and (b) ionic contribution to the dielectric function of pristine Cs$_2$AgInCl$_6$ and different alloyed compounds.}
	\label{2}	
\end{figure} 

In Fig.~\ref{2}(a), we have shown the electronic contribution of the dielectric function calculated using G$_0$W$_0$@HSE06 and in Fig.~\ref{2}(b), we have plotted the ionic contribution to the dielectric function computed using DFPT approach. The value of $\epsilon_\textrm{$\infty$}$ and $\epsilon_\textrm{static}$ for pristine Cs$_2$AgInCl$_6$ double perovskite are 2.05 and 9.63, respectively, which are well in agreement with the obtained values in previous reports~\cite{doi:10.1063/5.0031336,manna2020lattice}. Also from Fig.~\ref{2}(b), we have observed that the value of $\epsilon_\textrm{static}$ for the alloyed compounds is less than the pristine Cs$_2$AgInCl$_6$, which implies that the ionic contribution to the dielectric constant in case of alloyed compounds is less prominent than the pristine Cs$_2$AgInCl$_6$. Now, using reduced mass (provided in Table~\ref{1}) and the dielectric constants, we have determined the upper and lower bounds of E$_\textrm{B}$ using Eq~\ref{1}, listed in Table~\ref{2}.
\begin{table}[htbp]
	\caption{Upper bound (E$_\textrm{Bu}$) and lower bound (E$_\textrm{Bl}$) on exciton binding energy E$_\textrm{B}$ for double perovskites.} 
	\begin{center}
		\begin{adjustbox}{width=0.6\textwidth}
			\begin{tabular}[c]{|c|c|c|c|c|} \hline
				\textbf{Compounds} & \textbf{$\epsilon_\textrm{$\infty$}$}  & \textbf{E$_\textrm{Bu}$ (eV)} &  \textbf{$\epsilon_\textrm{static}$} & \textbf{E$_\textrm{Bl}$ (eV)}\\ \hline
				Cs$_2$AgInCl$_6$  & 2.05 & 0.74 & 9.63 & 0.03 \\ \hline
				Cs$_2$AgInBr$_{0.04}$Cl$_{5.96}$  & 2.06 & 0.67 & 6.90 & 0.06 \\ \hline
				Cs$_2$AgInI$_{0.04}$Cl$_{5.96}$   & 2.09 & 0.59 & 5.25 & 0.09 \\ \hline
				Cs$_2$Au$_{0.25}$Ag$_{0.75}$InCl$_6$   & 2.10 & 0.58 & 5.64 &  0.08\\ \hline
				Cs$_2$Cu$_{0.25}$Ag$_{0.75}$InCl$_6$ & 2.07 & 0.70 & 3.71 & 0.21\\ \hline
				Cs$_2$AgGa$_{0.25}$In$_{0.75}$Cl$_6$  & 2.05 & 0.68 & 4.71 & 0.12\\ \hline
				Cs$_2$AgIr$_{0.25}$In$_{0.75}$Cl$_6$  & 2.09 & 0.87 & 3.72 & 0.27 \\ \hline
				Cs$_2$Zn$_{0.50}$Ag$_{0.75}$In$_{0.75}$Cl$_6$ & 2.05 & 0.68 & 4.01 & 0.17 \\ \hline
			\end{tabular}
		\end{adjustbox}
		\label{Table2}
	\end{center}
\end{table}

\begin{figure}[h]
	\centering
	\includegraphics[width=0.6\textwidth]{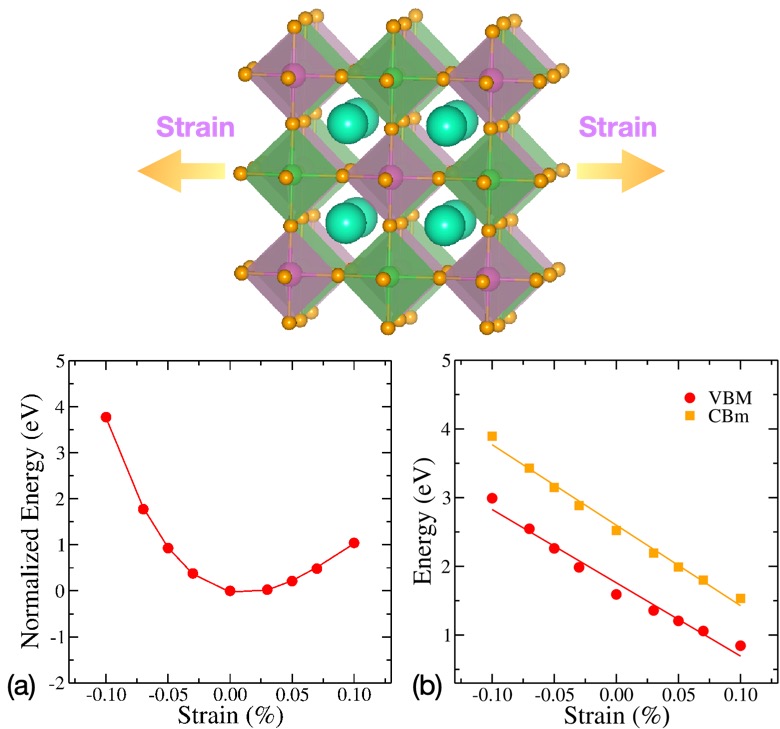}
	\caption{(a) Relationship between total energy and the applied strain along the transport direction (z-axis) (b) Band energy shifts of VBM and CBm under uniaxial strain for Cs$_2$AgInCl$_6$.}
	\label{3}	
\end{figure}

By knowing the exciton binding energy, dielectric function, and reduced mass, we can calculate the various excitonic parameters such as exciton radius (r$_{\textrm{exc}}$) and the probability of a wavefunction ($|\phi_\textrm{n}(0)|^2$) at zero charge separation  (given in Table~\ref{3}). The exciton radius ($\textrm{r}_\textrm{exc}$) is calculated as follows:
\begin{equation}
	\textrm{r}_\textrm{exc}=\frac{\textrm{m}_0}{\mu}\varepsilon_\textrm{eff}\,\textrm{n}^2\textrm{r}_\textrm{Ry}
\end{equation}
where $\textrm{m}_0$ is the free electron mass, $\mu$ is the reduced mass, $\varepsilon_\textrm{eff}$ is the effective dielectric constant (here, the electronic dielectric constant has been taken), n is the exciton energy level (n=1 provides the smallest exciton radius) and $\textrm{r}_\textrm{Ry}=0.0529$ nm is the Bohr radius. For electron-hole pair at zero seperation, the exciton lifetime ($\tau$) is inversely proportional to the probability of a wavefunction ($|\phi_\textrm{n}(0)|^2$).
The $|\phi_\textrm{n}(0)|^2$ is determined as follows:
\begin{equation}
	|\phi_\textrm{n}(0)|^2=\frac{1}{\pi(\textrm{r}_\textrm{exc})^3\textrm{n}^3}
\end{equation}
\begin{table}[h!]
	\caption{Excitonic parameters of Cs$_2$AgInCl$_6$ and different alloyed compounds.} 
	\begin{center}
		\begin{tabular}[c]{|c|c|c|} \hline
			\textbf{Compounds} & \textbf{r$_{\textrm{exc}}$ (nm)}  &   \textbf{|$\phi$$_\textrm{n}$(0)|$^2$ (10$^{27}$m$^{-3}$)} \\ \hline
			Cs$_2$AgInCl$_6$  & 0.47 & 2.96  \\ \hline
			Cs$_2$AgInBr$_{0.04}$Cl$_{5.96}$   & 0.53 & 2.18  \\ \hline
			Cs$_2$AgInI$_{0.04}$Cl$_{5.96}$  & 0.59 & 1.54  \\ \hline
			Cs$_2$Au$_{0.25}$Ag$_{0.75}$InCl$_6$   & 0.56 & 1.81  \\ \hline
			Cs$_2$Cu$_{0.25}$Ag$_{0.75}$InCl$_6$ & 0.50 & 2.54 \\ \hline
			Cs$_2$AgGa$_{0.25}$In$_{0.75}$Cl$_6$ & 0.53 & 2.13 \\ \hline
			Cs$_2$AgIr$_{0.25}$In$_{0.75}$Cl$_6$  & 0.49 & 2.87\\ \hline
			Cs$_2$Zn$_{0.50}$Ag$_{0.75}$In$_{0.75}$Cl$_6$ & 0.51 & 2.39  \\ \hline
		\end{tabular}
		\label{Table3}
	\end{center}
\end{table}

From Table~\ref{Table3}, it is observed that the alloyed compounds have longer exciton lifetime than the pristine Cs$_2$AgInCl$_6$ double perovskite. The longer exciton lifetime corresponds to lower recombination, which leads to higher quantum yield and conversion efficiency~\cite{selig2016excitonic}. Hence, the different alloyed compounds, having longer exciton lifetime than the pristine Cs$_2$AgInCl$_6$, are good for solar cells and photovoltaic devices. 

In order to calculate the charge carrier mobility, we have used deformation potential model proposed by Bardeen and Shockley~\cite{bardeen1950deformation}. According to this model, the charge carrier mobility ($\mu$) for 3D materials can be expressed as:
\begin{equation}
	\begin{split}
		\mu = \frac{2\sqrt{2\pi}e\hbar^4\textrm{C}_{\textrm{3D}}}{3(k_\textrm{B}\textrm{T})^{3/2}(\textrm{m}^*)^{5/2}\textrm{E}_{l}^2}
		\label{eq2}
	\end{split}
\end{equation}
where $k_\textrm{B}$ is the Boltzmann constant, T is the temperature, and \textit{e} is the elementary charge of electron (for more details, see section II of SI). We can see from Eq.~\ref{eq2}, the effective mass has a significant effect on the carrier mobility. Due to the intrinsic cubic structure, the pristine Cs$_2$AgInCl$_6$ and different alloyed compounds show an isotropic transport character. From Table~\ref{1}, it can be seen that the hole effective mass is nearly 3 times larger than that of the electron, which implies that electron transport ability is better than hole transport ability. 
\begin{table*}
	\caption{Calculated deformation potential constant ($\textrm{E}_{l}$), 3D elastic constant ($\textrm{C}_{\textrm{3D}}$), and carrier mobility ($\mu$) for electrons and holes along the transport direction (z-axis) of different configurations at 300 K.} 
	\begin{center}
			\begin{adjustbox}{width=0.85\textwidth}
		\begin{tabular}[c]{|c|c|c|c|c|} \hline
			\textbf{Compounds} & \textbf{Carrier type}  & \textbf{E$_l$ (eV)} &  \textbf{C$_{3\textrm{D}}$ (Nm$^{-2}$)} & \textbf{$\mu$ (cm$^2$V$^{-1}$s$^{-1}$)}\\ \hline
			\multirow{2}{*}{Cs$_2$AgInCl$_6$} & electron  & 11.71 & 67.48 & 665.38 \\ \cline{2-5}
			&  hole  & 10.66 & 67.48 & 30.68 \\ \hline			\multirow{2}{*}{Cs$_2$AgInBr$_{0.04}$Cl$_{5.96}$}  & electron  & 11.69 & 70.81 & 730.98 \\ \cline{2-5}
			& hole  & 11.26 & 70.81 & 59.21 \\ \hline
			\multirow{2}{*}{Cs$_2$AgInI$_{0.04}$Cl$_{5.96}$}  & electron  & 11.83 & 76.72 & 965.19 \\ \cline{2-5}
			& hole  & 11.91 & 76.72 & 90.05 \\ \hline
			\multirow{2}{*}{Cs$_2$Au$_{0.25}$Ag$_{0.75}$InCl$_6$}  & electron  & 11.57 & 67.25 & 909.88 \\ \cline{2-5}
			& hole  & 10.69 & 67.25 & 63.79 \\ \hline
			\multirow{2}{*}{Cs$_2$Cu$_{0.25}$Ag$_{0.75}$InCl$_6$}  & electron  & 11.54 & 62.43 & 765.15\\ \cline{2-5}
			& hole  & 9.78 & 62.43 & 31.98\\ \hline
			\multirow{2}{*}{Cs$_2$AgGa$_{0.25}$In$_{0.75}$Cl$_6$}  & electron  & 11.47 & 66.75 & 804.62\\ \cline{2-5}
			& hole  & 10.36 & 66.75 & 59.06\\ \hline
			\multirow{2}{*}{Cs$_2$AgIr$_{0.25}$In$_{0.75}$Cl$_6$}  & electron  & 11.53 & 72.08 & 673.36 \\ \cline{2-5}
			& hole  & 8.57 & 72.08 & 88.10 \\ \hline
			\multirow{2}{*}{Cs$_2$Zn$_{0.50}$Ag$_{0.75}$In$_{0.75}$Cl$_6$} & electron  & 11.38 & 58.97 & 672.35 \\ \cline{2-5}
			& hole  & 10.31 & 58.97 & 53.86  \\ \hline			
		\end{tabular}
			\end{adjustbox}
		\label{4}
	\end{center}
\end{table*}
Further, along with effective mass, the carrier mobility also depends on 3D elastic constant $\textrm{C}_{\textrm{3D}}$ and deformation potential $\textrm{E}_{l}$. In order to compute $\textrm{C}_{\textrm{3D}}$ and $\textrm{E}_{l}$, uniaxial strain has been applied along the lattice direction (z-axis). The total energies and positions of the VBM and CBm are calculated with respect to the uniaxial strain using PBE xc functional. These values are also calculated using HSE06 xc functional for Cs$_2$AgInCl$_6$ and Cs$_2$Cu$_{0.25}$Ag$_{0.75}$InCl$_6$, which are comparable with PBE xc functional (see section III of SI for comparison). Therefore, we have used PBE xc functional as it is computationally more cost effective. Now, $\textrm{C}_{\textrm{3D}}$ can be obtained by fitting the energy vs strain curve and $\textrm{E}_{l}$ is proportional to the band edge shift induced by applied strain along the transport direction (see Fig.~\ref{3} for Cs$_2$AgInCl$_6$). Subsequently, based on the obtained values of m$^*$, $\textrm{C}_{\textrm{3D}}$ and $\textrm{E}_{l}$, the electron mobility comes out to be 665.38 cm$^2$V$^{-1}$s$^{-1}$ and the hole mobility is 30.68 cm$^2$V$^{-1}$s$^{-1}$ for Cs$_2$AgInCl$_6$. The obtained hole mobility is quite well in agreement with the previous finding~\cite{doi:10.1021/acs.jpclett.9b00134}. Note that, the electron mobility is about 20 times higher than that of the hole mobility. Further, for the alloyed compounds, we have observed increment in carrier mobilities in comparison to pristine Cs$_2$AgInCl$_6$ (see Table~\ref{4}). This demonstrates excellent photovoltaic properties for the alloyed compounds than the pristine (For more details of $\textrm{C}_{\textrm{3D}}$ and $\textrm{E}_{l}$ for alloyed compounds, see section IV in SI).
\begin{table*}[h]
	\caption{Polaron parameters corresponding to electrons in pristine Cs$_2$AgInCl$_6$ and different alloyed compounds.} 
	\begin{center}
		\begin{tabular}[c]{|c|c|c|c|c|c|c|c|c|} \hline
			\textbf{Compounds}  & \textbf{$\omega_{\textrm{LO}}$ (THz)} & \textbf{$\theta_\textrm{D}$ ($\textrm{K}$)} & \textbf{$\alpha_\textrm{e}$} &  \textbf{$\textrm{m}_\textrm{P}$/$\textrm{m}^*$} &  \textbf{l$_{\textrm{P}}$ (\AA)} & \textbf{$\mu_{\textrm{P}}$ (cm$^2$V$^{-1}$s$^{-1}$)}\\ \hline
			Cs$_2$AgInCl$_6$  & 4.49 & 215 &  5.59 & 2.71 & 28.73 & 8.54\\ \hline
			Cs$_2$AgInBr$_{0.04}$Cl$_{5.96}$  & 5.86 & 281 &  4.26 & 2.16 & 40.48 & 13.44\\ \hline
			Cs$_2$AgInI$_{0.04}$Cl$_{5.96}$   & 5.85 & 280 &  3.48 & 1.88 & 45.09 & 21.42\\ \hline
			Cs$_2$Au$_{0.25}$Ag$_{0.75}$InCl$_6$  & 5.60 & 268 &  3.62 & 1.93 & 42.75 & 21.17\\ \hline
			Cs$_2$Cu$_{0.25}$Ag$_{0.75}$InCl$_6$ & 5.69 & 273 &  2.66 & 1.62 & 50.75 & 32.53\\ \hline
			Cs$_2$AgGa$_{0.25}$In$_{0.75}$Cl$_6$   & 5.36 & 257 &  3.54 & 1.90 & 41.85 & 20.81\\ \hline
			Cs$_2$AgIr$_{0.25}$In$_{0.75}$Cl$_6$  & 5.12 & 245 &  2.86 & 1.68 & 45.24 & 28.35\\ \hline
			Cs$_2$Zn$_{0.50}$Ag$_{0.75}$In$_{0.75}$Cl$_6$  & 3.07 & 147 &  4.13 & 2.11 & 25.56 & 21.60 \\ \hline			
		\end{tabular}
		\label{Table5}
	\end{center}
\end{table*}\\ 

The influence of electron-phonon coupling on physical/chemical properties of a material remains an alluring paradox. It has been reported that in case of polar semiconductors (e.g., halide perovskites), the interaction of carriers with the macroscopic electric field generated by longitudinal optical (LO) phonons, known as the Fr\"{o}hlich interaction is the dominating scattering mechanism near room temperature~\cite{frohlich1954electrons,herz2017charge}. In order to find the reason behind the discrepancy in mobility between theory and experiment, Fr\"{o}hlich's polaron model is observed to be a remarkable model. Using this model, here we have investigated the electron-phonon coupling in Cs$_2$AgInCl$_6$ and different alloyed compounds and estimated an upper limit on the related charge carrier mobilities.
The dimensionless Fr\"{o}hlich parameter $\alpha$ of dielectric electron-phonon coupling which can be seen as a comparative measure of Fr\"{o}hlich coupling strength is given as
\begin{equation}
	\begin{split}
		\alpha = \frac{1}{4\pi\epsilon_0}\frac{1}{2}\left(\frac{1}{\epsilon_\infty}-\frac{1}{\epsilon_\textrm{static}}\right)\frac{{e}^2}{\hbar\omega_{\textrm{LO}}}\left({\frac{2\textrm{m}^*\omega_{\textrm{LO}}}{\hbar}}\right)^{1/2}
		\label{eq1}
	\end{split}
\end{equation}
This parameter $\alpha$ is fully defined by the material specific properties, specifically, optical ($\epsilon_{\infty}$) and static ($\epsilon_{\textrm{static}}$) dielectric constants, the carrier effective mass ($\textrm{m}^*$), and a characteristic phonon angular frequency ($\omega_{\textrm{LO}}$). For a system with multiple phonon branches, an average LO frequency can be calculated by considering all the infrared active optical phonon branches and taking a spectral average of them~\cite{hellwarth1999mobility}. $\epsilon_0$ is the permittivity of free space. We observe that due to an increase in high frequency dielectric constant, the Fr\"{o}hlich scattering strength in alloyed compounds is reduced (see Table~\ref{Table5}). Note that, the value of $\alpha$ remains in the moderate strength range, and comparable to the hybrid halide perovskite series~\cite{doi:10.1063/5.0044146}. Further, the Debye temperature ($\theta_\textrm{D}$) of effective LO frequency comes out to be 215 K for pristine Cs$_2$AgInCl$_6$. Similarly, for the different alloyed systems as well, the Debye temperature is well below the room temperature and hence, suggests dominant polaronic contribution to limit carrier mobility near room temperature.

From the above discussion, we have found that these materials possess large dielectric electron-phonon coupling. Polarons are dressed quasiparticles, which are formed due to the interaction of electron and hole with the lattice. Within Fr\"{o}hlich's polaron theory, as extended by Feynman, the polaron mass renormalisation for polaron, ${\textrm{m}_\textrm{p}}$~\cite{feynman1955slow} can be calculated as
\begin{equation}
	\begin{split}
		{\textrm{m}_{\textrm{P}}} = \textrm{m}^*\left(1+\frac{\alpha}{6} +  \frac{\alpha^2}{40}\right)
		\label{eq4}
	\end{split}
\end{equation}
Further, the polaron radius can be calculated as
\begin{equation}
	\begin{split}
		\textrm{l}_\textrm{P} = \sqrt{\frac{h}{2\textrm{c}\textrm{m}^*\omega_{\textrm{LO}}}}
		\label{eq5}
	\end{split}
\end{equation} 
where, c is the speed of light. Our calculated polaron mass renormalisation and the corresponding polaron radius are listed in Table~\ref{Table5}. Notably, these parameters can be used to estimate an upper limit on charge carrier mobilities under the assumption that carriers are interacting only with the optical phonons. Now, the polaron mobility according to the Hellwarth polaron model~\cite{hellwarth1999mobility} is defined as follows: 
\begin{equation}
	\begin{split}
		\mu_{\textrm{P}} = \frac{(3\sqrt{\pi}e)}{2\pi \textrm{c}\omega_{\textrm{LO}}\textrm{m}^*\alpha} \frac{\textrm{sinh}{(\beta/2)}}{\beta^{5/2}}\frac{w^3}{v^3}\frac{1}{\textit{K}}
		\label{eq5}
	\end{split}
\end{equation}
where, $\beta$ = hc$\omega_{\textrm{LO}}$/$k_{\textrm{B}}$T, $\textit{w}$ and $\textit{v}$ correspond to temperature dependent variational parameters. $\textit{K}$ is a function of $\textit{v}$, $\textit{w}$, and $\beta$. Here, we have calculated $\textit{w}$ and $\textit{v}$ by minimizing the free polaron energy (see section V in SI for details). The obtained upper limits on the charge carrier mobilities at room temperature are listed in Table~\ref{Table5}.
From Table~\ref{Table5}, one can find that in case of pristine Cs$_2$AgInCl$_6$ and different alloyed compounds, large polarons [$\textrm{l}_\textrm{P}$ $>>$ lattice parameter] will be formed. This leads to increased carrier lifetime due to the reduction in carrier-carrier and carrier-defect scatterings~\cite{zheng2019large}. Large polarons can be effectively seen as quasicharge carriers possessing higher effective masses than the original carriers, thus practically reducing the carrier mobility. In case of pristine Cs$_2$AgInCl$_6$, our estimated polaron mobility is 8.54 cm$^2$V$^{-1}$s$^{-1}$, which suggests the dominant role of electron-phonon coupling here. We have also found polaronic mass to be 2.71 times of the effective mass of electron, which confirms the increased carrier lattice interaction. Further, in case of alloyed compounds, we have observed substantial reduction in electronic effective mass and polaronic mass renormalization (see Table~\ref{Table5}). As such, the limit to the electron mobility increases almost 2-4 times, in case of alloyed compounds. We have discussed the hole mobility as well (for reference, see SI section VI).
From the above discussion, we infer that the reduction in effective mass and increase in dielectric constant will largely improve the carrier mobility and ensure a better screening against ionized impurities in different alloyed compounds compared to pristine Cs$_2$AgInCl$_6$. This shows the higher overall carrier mobility in different alloyed compounds.

In summary, using first-principles calculation, the excitonic, and polaronic properties of pristine Cs$_2$AgInCl$_6$ and different alloyed double perovskites are investigated systematically. Initially, using Wannier-Mott approach, we have computed the upper and lower bound of exciton binding energy. After that, from the exciton lifetime calculation, we have found that the alloyed compounds have longer exciton lifetime than the pristine Cs$_2$AgInCl$_6$. Further, the deformation potential approach reveals substantial increase in the charge carrier mobility on alloying. Using the Feynman polaron model, we have discussed the carrier-lattice interaction in these materials. Careful analysis of carrier effective mass, electronic and ionic dielectric constant and effective LO frequency help us to address the electron-phonon coupling. The detailed theoretical investigation presented in this work will surely help the future studies to improve the overall performance of double perovskites.
\section{Computational Methods}
The density functional theory (DFT)~\cite{PhysRev.136.B864,PhysRev.140.A1133} calculations have been performed using the Vienna \textit{ab initio} simulation package (VASP)~\cite{KRESSE199615,PhysRevB.59.1758}. The ion-electron interactions in all the elemental constituents are described using projector-augmented wave (PAW) potentials~\cite{PhysRevB.50.17953,PhysRevB.59.1758}. The double perovskite Cs$_2$AgInCl$_6$ has a cubic structure with space group \textit{Fm$\bar{3}$m}. The corresponding sublattice is composed of alternate octahedra of InCl$_6$ and AgCl$_6$ as shown in Fig.~\ref{1}(a). All the structures are optimized (the atomic positions are relaxed) using generalized gradient approximation of PBE~\cite{PhysRevLett.77.3865} as the exchange-correlation (xc) functional until the forces are smaller than 0.001 eV/\AA. The electronic self consistency loop convergence is set to 0.01 meV, and the kinetic energy cutoff is set to 600 eV for plane wave basis set expansion. A $\textit{k}$-grid of $4\times4\times4$ is used for Brillouin zone integration, which is generated using Monkhorst-Pack~\cite{PhysRevB.13.5188} scheme. Advanced hybrid xc functional HSE06~\cite{doi:10.1063/1.2404663} is used for the better estimation of band gap. For determination of optical properties, single shot GW (G$_0$W$_0$)~\cite{PhysRev.139.A796,PhysRevLett.55.1418} calculations have been performed on top of the orbitals obtained from HSE06 xc functional. The polarizability calculations are carried out on a grid of 50 frequency points. The number of bands is set to four times the number of occupied orbitals. Note that spin orbit coupling was not taken into consideration here because it negligibly affects the electronic structures of Ag/In halide double perovskites~\cite{C7MH00239D,doi:10.1021/acs.chemmater.9b00116}. The ionic contribution to dielectric function has been calculated using density functional perturbation theory (DFPT)~\cite{gajdovs2006linear} with $7\times7\times7$ $\textit{k}$-grid. Carrier mobilities are simulated via implementing a temperature dependent Feynman polaron model~\cite{PhysRev.127.1004}. A detailed description of the model along with the necessary parameters needed can be found in Ref~\cite{PhysRev.127.1004,PhysRevB.96.195202,PhysRev.130.1364}.
\begin{acknowledgement}
MJ acknowledges CSIR, India, for the senior research fellowship [grant no. 09/086(1344)/2018-EMR-I]. MK acknowledges CSIR, India, for the senior research fellowship [grant no. 09/086(1292)/2017-EMR-I]. PB acknowledges UGC, India, for the senior research fellowship [1392/(CSIR-UGC NET JUNE 2018)]. SB acknowledges the financial support from SERB under core research grant (grant no. CRG/2019/000647). We acknowledge the High Performance Computing (HPC) facility at IIT Delhi for computational resources.
\end{acknowledgement}
\begin{suppinfo}
Electronic bandstructure of different alloyed compounds using G$_0$W$_0$@HSE06; Free charge carrier mobility using deformation potential model; Comparison of charge carrier mobility using PBE and HSE06 exchange-correlation (xc) functional; Deformation potential constant ($\textrm{E}_{l}$) and 3D elastic constant ($\textrm{C}_{\textrm{3D}}$) for alloyed compounds; Polaron mobility using Feynman polaron model; Polaron parameters for holes.
	
\end{suppinfo}
\bibliography{achemsodemo.bib}
\end{document}


\begin{center}
{\Large \bf Supplemental Material}\\ 
\end{center}
\begin{enumerate}[\bf I.]
	
\item Electronic bandstructure of different alloyed compounds using G$_0$W$_0$@HSE06.	
\item Free charge carrier mobility using deformation potential model.
\item Comparison of charge carrier mobility using PBE and HSE06 exchange-correlation (xc) functional.
\item  Deformation potential constant ($\textrm{E}_{l}$) and 3D elastic constant ($\textrm{C}_{\textrm{3D}}$) for alloyed compounds. 	
\item Polaron mobility using Feynman polaron model.
\item Polaron parameters for holes.

\end{enumerate}
\vspace*{12pt}
\clearpage

\section{Electronic bandstructure of different alloyed compounds using G$_0$W$_0$@HSE06}
\begin{figure}[h]
	\centering
	\includegraphics[width=1.05\textwidth]{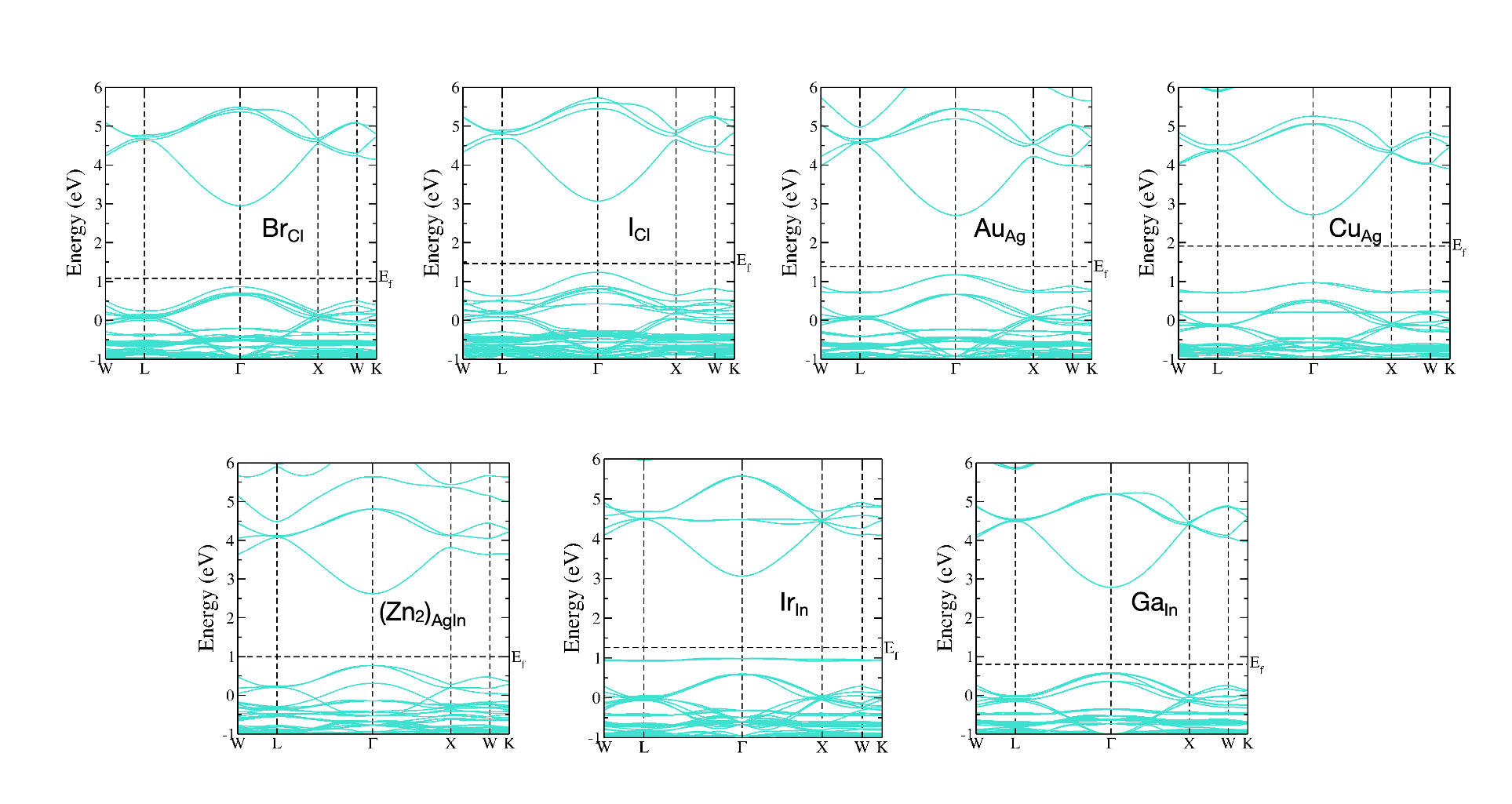}
	\caption{Electronic bandstructure of different alloyed compounds. E$_\textrm{f}$ is the fermi energy level.}
	\label{1}	
\end{figure} 

\section{Free charge carrier mobility using deformation potential model}
According to deformation potential model~\cite{bardeen1950deformation}, the carrier mobility is defined as: 
\begin{equation}
	\begin{split}
		\mu = \frac{2\sqrt{2\pi}e\hbar^4\textrm{C}_{\textrm{3D}}}{3(k_\textrm{B}\textrm{T})^{3/2}(\textrm{m}^*)^{5/2}\textrm{E}_{l}^2}
		\label{eq1}
	\end{split}
\end{equation}
where $k_{\textrm{B}}$,  T,  and m$^*$ correspond to the Boltzmann constant, temperature, and effective mass, respectively. Here, we have performed the calculations at 300 K temperature. In the above equation, E$_l$ corresponds to the deformation potential constant along z-direction for electron and hole. It is given by:
\begin{equation}
	\begin{split}
		\textrm{E}_l = \frac{\Delta \textrm{E}_\textrm{i}}{\Delta \textrm{l}/\textrm{l}_0}
		\label{eq2}
	\end{split}
\end{equation}
where, $\Delta \textrm{E}_\textrm{i}$ denotes the change in the energy or shift in the position of VBM or CBm under uniaxial strain along the z-direction. l$_0$ is the lattice constant along the transport direction (z-axis). $\Delta \textrm{l}$ denotes the change or deformation in the lattice constant l$_0$ on application of uniaxial strain.  The elastic modulus C$_{3\textrm{D}}$ is computed using the relation (E - E$_0$)/V$_0$ = C($\Delta \textrm{l}/\textrm{l}_0)^2/2$, where E$_0$ and E are the total energies of undeformed and deformed system, respectively. V$_0$ represents the equilibrium volume of the system. Note, herein, we have chosen strain range from -1.0\% to +1.0\% to obtain the fitted values of C$_{3\textrm{D}}$ and E$_l$.

\section{Comparison of charge carrier mobility using PBE and HSE06 exchange-correlation (xc) functional}
\begin{table*}[h]
	\caption{Calculated deformation potential constant ($\textrm{E}_{l}$), 3D elastic constant ($\textrm{C}_{\textrm{3D}}$), carrier mobility ($\mu$) for  holes along the transport direction (z-axis) of pristine Cs$_2$AgInCl$_6$ and  	Cs$_2$Cu$_{0.25}$Ag$_{0.75}$InCl$_6$ at 300 K.} 
	\begin{center}
		\begin{adjustbox}{width=1.00\textwidth}
		\begin{tabular}[c]{|c|c|c|c|c|c|c|} \hline
			& \multicolumn{3}{c|}{$\textbf{PBE}$} 	& \multicolumn{3}{c|}{$\textbf{HSE06}$} \\ \hline
			\textbf{Compounds}  & \textbf{E$_l$ (eV)} & \textbf{C$_{3\textrm{D}}$ (Nm$^{-2}$)} & \textbf{$\mu$ (cm$^2$V$^{-1}$s$^{-1}$)} & \textbf{E$_l$ (eV)} & \textbf{C$_{3\textrm{D}}$ (Nm$^{-2}$)} & \textbf{$\mu$ (cm$^2$V$^{-1}$s$^{-1}$)} \\ \hline
			Cs$_2$AgInCl$_6$  & 10.66 & 67.48 &  30.68  & 10.81 & 70.54 &  31.23 \\ \hline
			Cs$_2$Cu$_{0.25}$Ag$_{0.75}$InCl$_6$ & 10.36 & 62.43 &  31.98 & 9.51 & 65.09 &  35.55 \\ \hline			
		\end{tabular}
		\end{adjustbox}
		\label{Table1}
	\end{center}
\end{table*}
As we can see from Table~\ref{Table1}, the carrier mobilities calculated using PBE~\cite{PhysRevLett.77.3865} and HSE06~\cite{doi:10.1063/1.2404663} xc functional are quite comparable. Therefore, we have proceeded with PBE functional as it is cost effective. 
\newpage
\section{Deformation potential constant ($\textrm{E}_{l}$) and 3D elastic constant ($\textrm{C}_{\textrm{3D}}$) for alloyed compounds}
\begin{figure}[h!]
	\centering
	\includegraphics[width=0.90\textwidth]{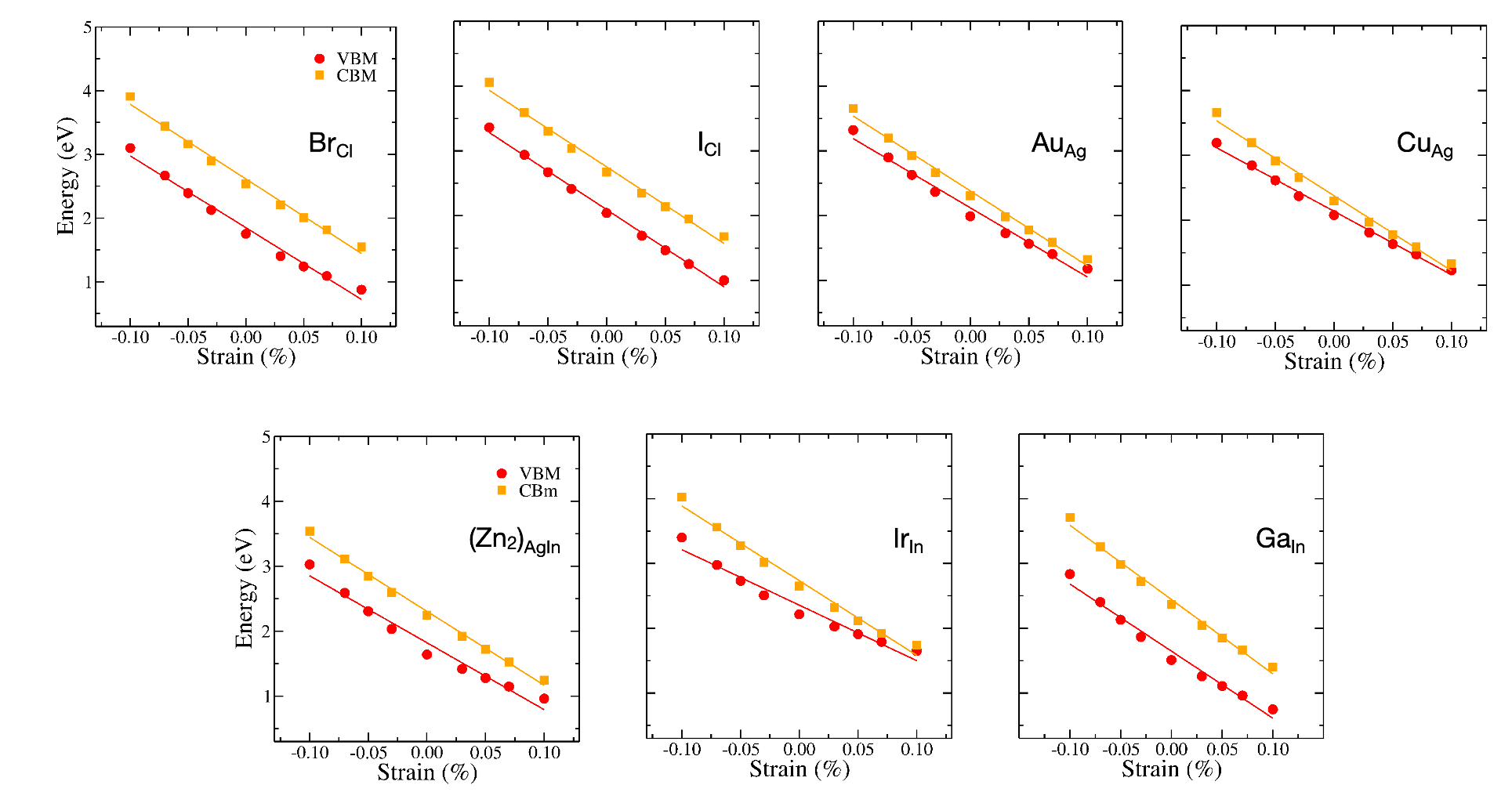}
	\caption{Energy shifts of CBm and VBM of different alloyed compounds under strain. The slopes of these lines are the corresponding deformation potential.}
	\label{2}	
\end{figure} 

\begin{figure}[h]
	\centering
	\includegraphics[width=0.90\textwidth]{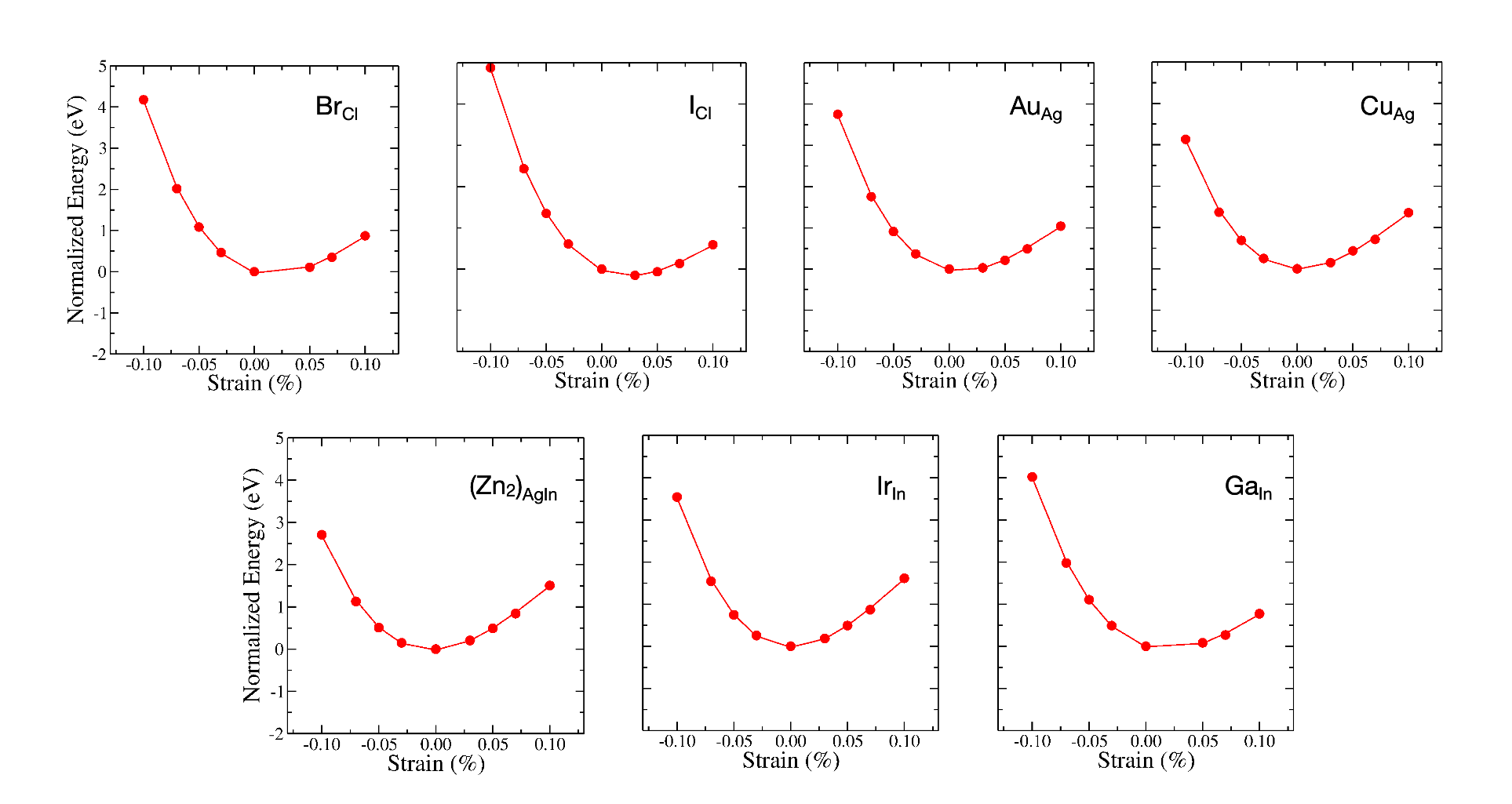}
	\caption{Relationship between total energy and the applied strain. The quadratic fitting of the data gives the elastic constant of different alloyed compounds.}
	\label{3}	
\end{figure}
\newpage

\section{Polaron mobility using Feynman polaron model}
Polaron mobility according to the Hellwarth polaron model~\cite{hellwarth1999mobility} is defined as follows: 
\begin{equation}
	\begin{split}
		\mu_{\textrm{P}} = \frac{(3\sqrt{\pi}e)}{2\pi \textrm{c}\omega_{\textrm{LO}}\textrm{m}^*\alpha} \frac{\textrm{sinh}{(\beta/2)}}{\beta^{5/2}}\frac{w^3}{v^3}\frac{1}{K}
		\label{eq3}
	\end{split}
\end{equation}
where, $\beta$ = hc$\omega_{\textrm{LO}}$/$k_{\textrm{B}}$T, \textit{e} is the electronic charge, m$^*$ is the effective mass of charge carrier, $\textit{w}$ and $\textit{v}$ correspond to temperature dependent variational parameters. $\textit{K}$ is a function of $\textit{v}$, $\textit{w}$, and $\beta$ defined as follows: 
\begin{equation}
	\begin{split}
		K(a,b) = \int_0^\infty du[u^2 + a^2 - b\textrm{cos}(vu)]^{-3/2}\textrm{cos}(u)
		\label{eq4}
	\end{split}
\end{equation}
In the above equation, a$^2$ and b can be calculated using following formulas: 
\begin{equation}
	\begin{split}
		a^2 = (\beta /2)^2 + \frac{(v^2 - w^2)}{w^2v} \beta coth(\beta v/2)
		\label{eq5}
	\end{split}
\end{equation} 
\begin{equation}
	\begin{split}
		b =  \frac{(v^2 - w^2)}{w^2v}\frac{\beta}{\textrm{sinh}(\beta v/2)}
		\label{eq6}
	\end{split}
\end{equation}  
\section{Polaron parameters for holes}
\begin{table*}[h]
	\caption{Polaron parameters corresponding to holes in pristine Cs$_2$AgInCl$_6$ and different alloyed compounds.} 
	\begin{center}
		\begin{adjustbox}{width=1.00\textwidth}
		\begin{tabular}[c]{|c|c|c|c|c|c|c|c|c|} \hline
			\textbf{Compounds}  & \textbf{$\omega_{\textrm{LO}}$ (THz)} & \textbf{$\theta_\textrm{D}$ ($\textrm{K}$)} & \textbf{$\alpha_\textrm{h}$} &  \textbf{$\textrm{m}_\textrm{P}$/$\textrm{m}^*$} &  \textbf{l$_{\textrm{P}}$ (\AA)} & \textbf{$\mu_{\textrm{P}}$ (cm$^2$V$^{-1}$s$^{-1}$)}\\ \hline
			Cs$_2$AgInCl$_6$  & 4.49 & 215 &  10.75 & 5.68 & 19.07 & 0.29\\ \hline
			Cs$_2$AgInBr$_{0.04}$Cl$_{5.96}$  & 5.86 & 281 &  7.21 & 3.50 & 29.86 & 1.22\\ \hline
			Cs$_2$AgInI$_{0.04}$Cl$_{5.96}$   & 5.85 & 280 &  5.59 & 2.71 & 34.79 & 3.03\\ \hline
			Cs$_2$Au$_{0.25}$Ag$_{0.75}$InCl$_6$  & 5.60 & 268 &  6.47 & 3.12 & 30.92 & 1.75\\ \hline
			Cs$_2$Cu$_{0.25}$Ag$_{0.75}$InCl$_6$ & 5.69 & 273 &  5.36 & 2.61 & 34.93 & 2.11\\ \hline
			Cs$_2$AgGa$_{0.25}$In$_{0.75}$Cl$_6$   & 5.36 & 257 &  6.43 & 3.10 & 30.10 & 1.67\\ \hline
			Cs$_2$AgIr$_{0.25}$In$_{0.75}$Cl$_6$  & 5.12 & 245 &  4.92 & 2.42 & 33.91 & 3.49\\ \hline
			Cs$_2$Zn$_{0.50}$Ag$_{0.75}$In$_{0.75}$Cl$_6$  & 3.07 & 147 &  7.11 & 3.44 & 19.09 & 2.37 \\ \hline			
		\end{tabular}
		\end{adjustbox}
		\label{Table2}
	\end{center}
\end{table*}
\bibliography{achemsodemo.bib}
\newpage